# Tight open knots

Piotr Pierański[1], Sylwester Przybył[1] and Andrzej Stasiak[2]

[1]Poznań University of Technology,
Nieszawska 13A, 60 965 Poznań, Poland
e-mail: Piotr.Pieranski@put.poznan.pl

[2] Laboratory of Ultrastructural Analysis, University of Lausanne, Switzerland

ABSTRACT

The most tight conformations of prime knots are found with the use of the SONO algorithm. Their curvature and torsion profiles are calculated. Symmetry of the knots is analysed. Connections with the physics of polymers are discussed.

## 1. Introduction

From the point of view of topology knots are closed, self-avoiding curves[1,2]. Tying a knot in practice we operate on a finite piece of a rope[3]. At the end of the knot tying procedure a compact knotted structure with the two loose ends is created. To fix the knot type we splice the ends of the rope (outside the knotted structure, of course); without cutting the rope, the knot type cannot be changed. The same knot type fixing effect is reached if instead of splicing the loose ends we pull them apart and attach to parallel walls. In what follows we shall refer to such structures as open knots.

Open knots are more common in nature than the closed knots. As indicated by de Gennes[4] such knots are spontaneously tied and untied on polymeric molecules. Their existence changes considerably macroscopic properties of the materials. In general, the topological aspects of the microscopic structure of polymeric materials prove to be an important issue of the physics of polymers[5,6,7].

As we know well from the everyday experience, knots tied on a piece of rope can be tightened by pulling apart its loose ends as much as possible. Obviously, there is always a limit to such a knot tightening process – a particular conformation of the knot is reached at which the loose ends cannot be pulled apart any more. We shall refer to such conformations as the tight open knots. As laboratory experiments prove, the final conformation of a tightened knot depends on two major factors: the initial conformation from which the knot tightening procedure starts and the physical parameters of the rope on which the knot is tied. The rope-rope friction coefficient, its elasticity constants etc. are to be taken into account. To get rid of such material dependent parameters we consider below knots tied on the perfect rope. The perfect rope has from the physical point of view somewhat contradictory properties:

  i. it is perfectly hard - to squeeze its circular cross-section into an ellipse the infinite energy is needed,
  ii. it is perfectly flexible - as long as none of its circular cross-sections overlap no energy is needed to bend it,
  iii. it is perfectly slippery - no energy is needed to tighten a knot tied on it.

Introduction of the notion of the perfect rope allows us to define better the subject of our study. In the same sense introduction of the notion of hard spheres clarified the formulation of the packing problems.

Take a piece of the perfect rope of length $L$ and tie a knot on it. Stretch the ends of the rope apart as much as possible and measure the distance $L'$ between them. Obviously, $L'\leq L$. The difference $L-L'$ can be seen as the length of the rope engaged within the knot. Dividing this value by the diameter $D$ of the used rope we obtain a dimensionless number $\Lambda$ known as the open thickness energy[8]. In the case of closed knots, in which the ends of the rope are spliced, $L'=0$. Knots in conformations for which $\Lambda$ reaches its global minimum are called ideal[9]. More detailed, rigorous analysis of the thickness energy functionals were performed by O'Hara, Simon, Rawdon and Millett[10]. The most tight open conformation of a knot of a given type, i.e. the conformation at which $\Lambda$ reaches its global minimum, is an interesting object of which very little is known. As suggested by the present authors[11], peaks within its curvature profile indicate places at which the real rope would be most prone to rupture. This hypothesis was verified on the trefoil knots tied on starch gel (spaghetti) filaments. Below we describe in detail the curvature and torsion profiles of tight open knots.

## 2. Open knots tied and tighened on the the perfect rope

Perfect ropes do not exist in nature. However, they are easily simulated in numerical experiments. The algorithm we used in the numerical experiments described below is SONO (Shrink-On-No-Overlaps) used previously in the search for the most tight (ideal) conformations of closed knots[12].

To demonstrate how SONO algorithm performs the knot tightening task we present results of a numerical experiment in which the initial conformation of the rope is so entangled, that at the first sight one cannot decide if it is knotted or not.

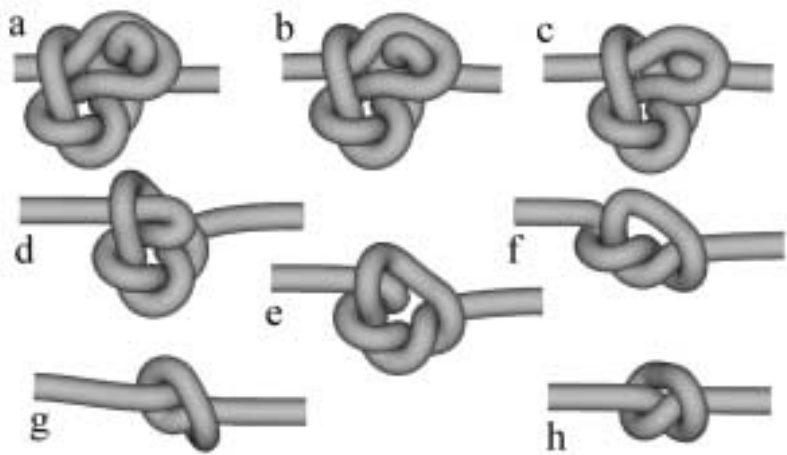

Fig. 1 Numerical simulation of axial shrinkage of the perfect rope entangled into a clumsy overhand knot (the open trefoil knot). Notice that nugatory crossings are easily removed while the entanglements due to knotting remain.

As seen in the figure, SONO algorithm performs the disentangling task without any problems. That the perfect rope is perfectly flexible and that there is no friction at its self-contact points is here of primary importance.

To compare the approximate values of $\Lambda$ we obtained in numerical experiments with a $\Lambda$ value known precisely, we considered also the case of the Hopf link for which the length of its open form would simply correspond to the minimal length of the unopened component threaded by the opened one. As easy to see $\Lambda$ equals in this case exactly $2\pi$. See Figure 2.

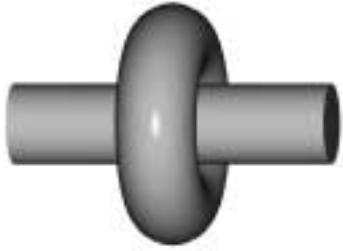

Figure 2. The ideal open Hopf link.

As laboratory experiments performed by Diao et al. indicated[8], $\Lambda$ varies with the knot type – different knots reduce the rope length in a different manner. Figure 3 presents the most tight conformations of the open trefoil ($3_1$) and figure-eight ($4_1$) knots tied and tightened by SONO on pieces of rope of identical length $L$.

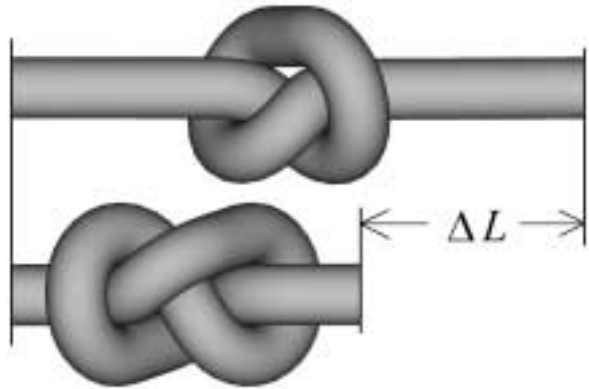

Fig. 3 The most tight open conformations of the trefoil ($3_1$) and figure-eight ($4_1$) knots tied on pieces of the rope of the same length.

Clearly, in accordance with laboratory experiments described in ref.8, the trefoil knot engages less rope than the figure-eight knot. The values of $\Lambda$ we obtained for these knots are, respectively, 10.1 and 13.7 ($\pm$ 0.05). Values provided by Diao et al.[8] are significantly larger: 10.35 and 14.65. The difference between results obtained in laboratory and numerical experiments exceeds errors specified by the authors. Most probably problems with friction encountered in experiments performed on real ropes do not allow one to enter the most tight conformations of the knots accessible in numerical experiments performed on perfect ropes. Table I shows the $\Lambda$ values we found for the most tight closed and open conformations of a few prime knots. To provide a natural $\Lambda$ unit, we included into the table also the rigorous value of $\Lambda$ known for the Hopf link  Interestingly enough, the difference of length of the most tight closed and open forms proves to be very close to 2π also for several other chiral knots. On the other hand, for achiral knots, $4_1$ and $6_3$, the differences are significantly different.

| Knot type | Open form | Closed form | Difference |
|---|---|---|---|
| Hopf link | 6.28 | 12.56 | 6.28 |
| $3_1$ | 10.1 | 16.33 | 6.23 |
| $4_1$ | 13.7 | 20.99 | 7.29 |
| $5_1$ | 17.3 | 23.55 | 6.25 |
| $5_2$ | 18.4 | 24.68 | 6.28 |
| $6_3$ | 20.7 | 28.88 | 8.18 |

Table I The normalized minimal length Λ of the most tight conformations of a few prime knots in their open and closed forms. Λ values for the closed forms of the knots were taken from ref.12. Rolfsen notation was used to indicate the knot types[13].

## 3. Symmetry of the curvature and torsion maps of the $3_1$ and $4_1$ open knots

Calculating the curvature and torsion of a smooth (differentiable twice) curve defined by analytical formulae is a trivial task. On the other hand, the determination of the curvature and torsion maps of the knots simulated in numerical experiments is extremely susceptible to the inaccuracies with which positions of the consecutive points of the simulated knot are given. To find the curvature, the first and second derivatives must be known. In the case of torsion, the third derivative is also needed. The discrete differentiation procedures reveal a considerable noise present within the curvature and torsion maps. To get rid of the noise, we took averages over a large set of the maps calculated in long runs in which the tightened knot was moving slowly there and back along the simulated rope. The slow oscillatory motion of the knot was introduced on purpose to minimise effects stemming from the discrete structure of the simulated rope. Figure 4 presents consecutive curvature maps registered in a short interval cut out from one of such runs. Averaging the instantaneous curvature maps within a reference frame which moves together with the knot we obtain curvature maps which are much better defined. Figure 5 presents such maps found for the most tight $3_1$ and $4_1$ knots. The knots were tied on the numerically simulated perfect rope consisting of 200 segments of equal length.

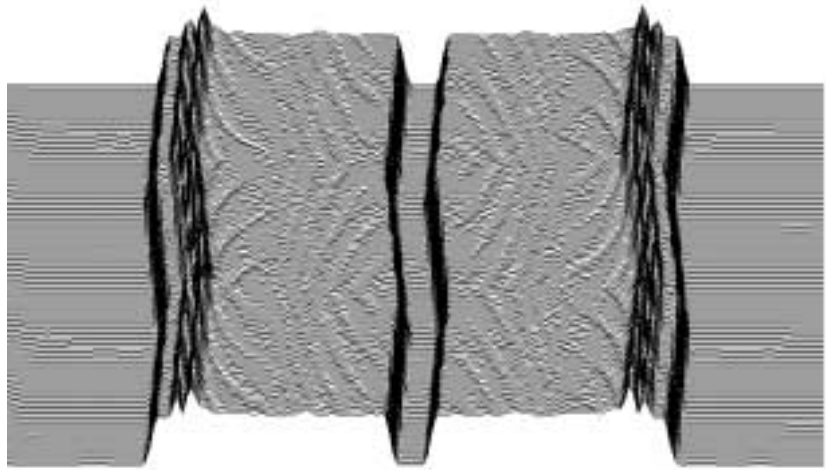

Fig.4 Curvature maps of the $3_1$ knot registered in a numerical experiment in which the tightened knot was slowly moving there and back along the simulated rope.

The plots display a few interesting features which we shall discuss below emphasising those of them, which according to us may survive in the limit of the most tight open conformations.

Let us start with the discussion of the curvature profiles. Within some accuracy limits, both profiles can be seen as consisting of two mirror symmetrical parts. We assume here that the external zero curvature regions are extended to infinity. The curvature profile of e.g. the $3_1$ knot seen from both ends of the knot looks almost identical. If by the curvature profile we understand the curvature $\kappa$ as a function of the arc length $l$, this symmetry property can be expressed as follows: in the middle of the knot there exists such a point $l_m$, that on the left and on the right of it curvature profiles are identical $\kappa(l_m+\varepsilon) = \kappa(l_m-\varepsilon)$.

In the case of the numerically simulated discretized knots the mirror symmetry is not exact. It seems to us, and we would like to express it as a conjecture, that in the case of the ideal open conformations of both $3_1$ and $4_1$ knots the mirror symmetry of the curvature profile should be exact. Mirror symmetry of the curvature profile reflects the twofold rotational symmetry of the open knot conformation. Let us note here that the point symmetry elements of the closed ideal conformations of the $3_1$ and $4_1$ knots are different.

One of the most distinct local features of the both curvature landscapes are the double peaks visible at the entrance to the knots. As described previously, the inner peak develops only at the final stage of the tightening process[11]. Laboratory experiments prove that at the points of high curvature the filaments, on which knots are tied, are most susceptible to breaking.

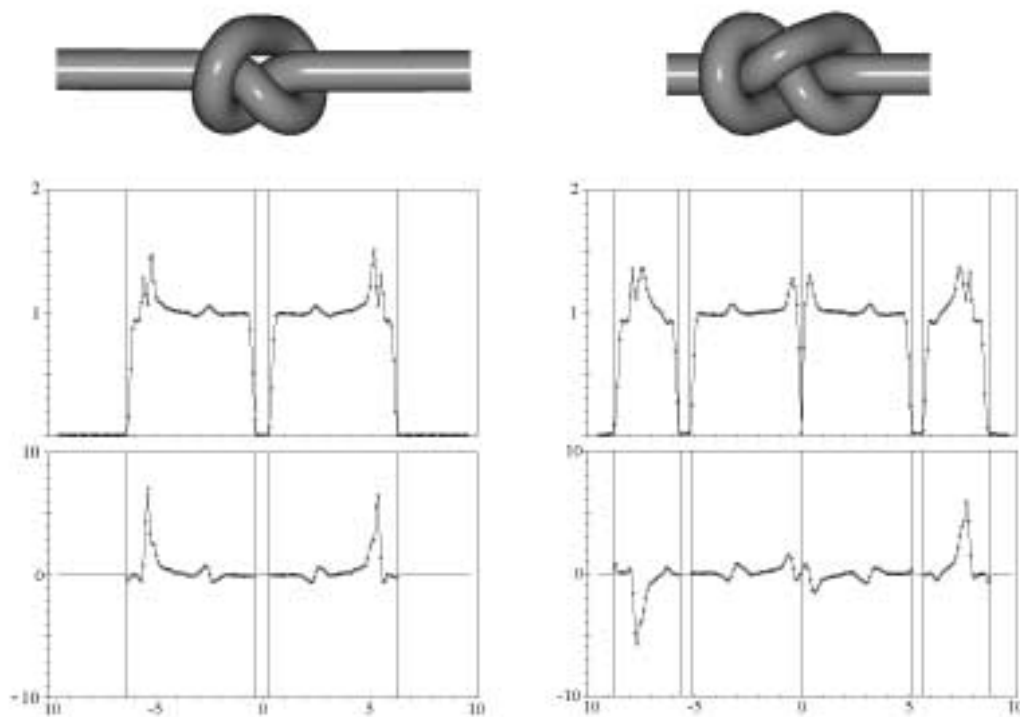

Fig. 5 The most tight open trefoil ($3_1$) and figure-eight ($4_1$) knots together with their maps of curvature and torsion.
 a) The trefoil knot shown in the figure (left) is of right-handed type and like other chiral knots cannot be converted into its mirror image.
 b) The map of curvature of the open trefoil knot shows a mirror symmetry (mirror plane is vertical and is in the centre of the map) thus, an identical map of curvature would be obtained for the left handed enantiomer of this knot. The curvature is normalised in respect to the diameter of tightly knotted tubes: value 1 is attained when the local radius of curvature corresponds to the diameter of the tube, value 2 would correspond to the radius of curvature being equal to the radius of the tube (points of sharp reversals).
 c) The map of torsion of the open trefoil knot shows also a mirror symmetry, thus, for left-handed enantiomer the map of torsion would be reversed along the vertical axis.
 d) Figure eight knot is achiral and is easily convertible into its mirror image (see Fig. 6).
 e) The map of curvature of the open figure eight knot shows perfect mirror symmetry and thus has no polarity.
 f) The map of torsion of the open figure eight knot shows a clear polarity - the passage from the left to right is different from this from the right to left. The total torsion is however zero as it is expected for an ideal form of achiral knot.

The maps of torsion show important differences between chiral $3_1$ knot and achiral $4_1$ knot. The map of torsion of the $3_1$ knot shows a mirror symmetry while the torsion map of the achiral knot $4_1$ shows a symmetry of different kind: in the middle of the knot there exists such a point $l_m$, that on the left and on the right of it torsion profiles are of identical magnitude but of an opposite sign $\kappa(l_m+\varepsilon) = -\kappa(l_m-\varepsilon)$. Thus, observing the torsion we can distinguish between the two ends of this knot. Taking into account curvature and torsion, which provide the complete descriptors of a given trajectory, we see that a time reversed travel through a chiral knot $3_1$ is indistinguishable from the original one. In contrast to that, in the achiral $4_1$ knot we can distinguish between time reversed travels – the signs of the torsion component appear during the travels in the reversed order. In the $4_1$ knot the time reversed travel is identical with the not reversed travel along the mirror image of the original knot (the mirror reflection changes left-handed regions into right-handed and contrary). As could be expected for achiral knots the total torsion cancels to zero while this of course is not the case for achiral knots. It may seem surprising that achiral knots $4_1$ are in fact polar while this is not the case for a chiral knot $3_1$. Interestingly, the ideal open configuration of $4_1$ knot is congruent with its mirror image. Figure 6.

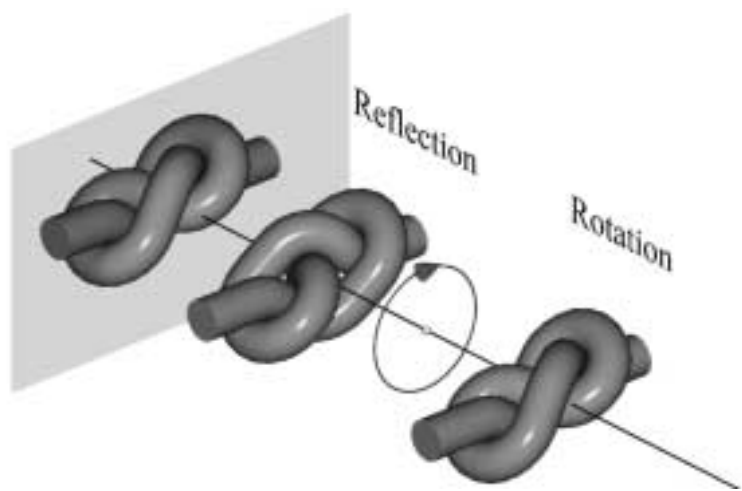

Fig.6 Congruency of the open $4_1$ knot with its mirror image. Rotation by $\pi/2$ is followed by a mirror reflection.

Although, by definition, all achiral knots can be continuously converted into their mirror images there is only a small subset of their configurations which upon rigid transformation are congruent with their mirror images[14].

**4. The problem of the local minima**

Several independent starting configurations of such simple knots as $3_1$ or $4_1$ were converted during our simulations to configurations essentially identical to those shown in Fig. 3. This was, however, not the case for more complicated knots. Fig.7 shows what happens when the most tight closed configuration of $5_2$ knot is opened in three different positions and the ends are pulled apart. Three different final configurations are obtained and to pass from one to another the string has to be loosened and loops have to be moved along the knot. Apparently the three different configurations constitute different local minima in the configuration space of tight open knots. Interestingly the nice symmetrical configuration shown in Fig. 7a causes biggest effective shortening of the string and is thus furthest from the global minimum. The configuration shown on Fig. 7c may in fact represent a global minimum for this knot as its effective shortening of the tube is smallest. This configuration does not show any symmetry.

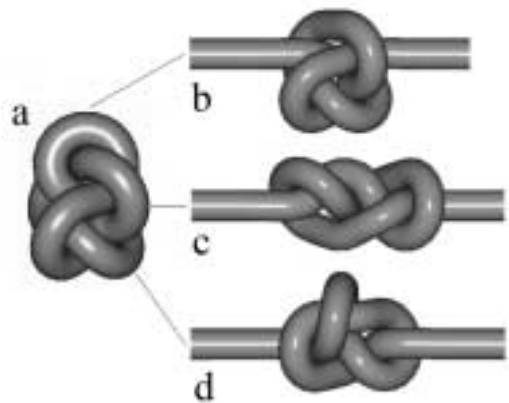

Fig. 7 Three different local minima in the configuration space of open knots $5_2$. (a) Knot $5_2$ in its most tight closed configuration. Three distinct sites 1, 2 and 3 were used to open the knot. (b), (c), (d) The local minima configurations obtained upon opening the closed knot in sites 1, 2 and 3 respectively. Notice that the symmetrical conformation 1 engages more of the rope length than the other two configurations. The conformation 3 represents presumably a global minimum and thus would corresponds to an ideal open configuration of $5_2$ knot.

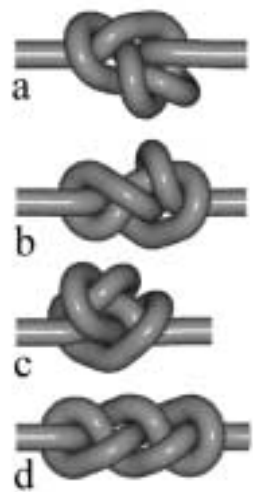

Fig. 8 Local minima conformations of the $6_3$ knot. Notice the symmetry of the conformation marked as (d). See text for its discussion.

To check if symmetrical configurations may constitute local minima in configuration space of more complex tight open knots we took the most tight closed configuration of achiral knot $6_3$ and opened it in different positions. See Fig. 8. Upon pulling apart the opened ends we noticed that one of the openings led to a nice regular form congruent with its mirror image. It seems to us that this form constitutes the global minimum.

**5. Discussion**

Perfect ropes do not exist in nature, but polymeric molecules are not far from being perfect. Put into the perpetual motion by thermal fluctuations they never get blocked by friction which plays such an important role in the macroscopic world allowing, for instance, to stop the huge mass of a docking ship with a rope tied in an appropriate manner (round turn and two half hitches) to the bollard[15]. Time averaged conformations of the open knots tied on polymeric molecules are very similar to the ideal open knots we considered above. In particular, the length of the molecular rope engaged within such knots should be directly related to the $\Lambda$ value we calculated. Properties of the knotted molecules are essentially different form the unknotted ones. For instance, their gel mobility coefficients are essentially different[16,17]. In general, polymeric materials in their phases in which the fraction of knotted molecules is considerable, should display interesting physical properties[4]. When the polymer technology will provide us with such materials is difficult to predict today. Certainly, the properties of a polymer, whose all molecules are tied into right-handed trefoil knots will be different from the polymer within which all the knots are left-handed.

**Acknowledgement**

We thank G. Dietler and J. Dubochet for helpful discussions. This work was carried out under projects: KBN 5PO3B01220 and SNF 31-61636.00.

[1] L. H. Kauffman, *Knots and Physics* (World Scientific Publishing Co., 1993).

[2] C. C. Adams, *The Knot Book* (W.H. Freeman and Company, New York, 1994).

[3] C. W. Ashley, *The Ashley Book of Knots* (Doubleday, New York, 1993)

[4] P.-G. de Gennes, Macromolecules, **17**, 703 (1984).

[5] M. D. Frank-Kamentskii and A. V. Vologodskii, Sov. Phys. Usp. **24** 679 (1981);

[6] A. Y. Grossberg, A. R. Khokhlov, *Statistical physics of macromolecules*, AIP Press, 1994.

[7] A. Y. Grossberg, A. Feigel and Y. Rabin, *Phys. Rev.* **A 54**, 6618-6622 (1996).

[8] Y. Diao, C. Ernst and E. J. Janse van Rensburg in *Ideal Knots*, eds. Stasiak, A., Katritch, V. and Kauffman, L. H. World Scientific, Singapore, 1998, p.52-69.

[9] V. Katritch, J. Bednar, D. Michoud, R. G. Sharein, J. Dubochet and A. Stasiak, Nature **384**, 142 -145 (1996); V. Katritch, W. K. Olson, P. Pierański, J. Dubochet and A. Stasiak, Nature **388**, 148 (1997).

[10] See chapters by: E. Rawdon; J. A. Calvo and K. C. Millett; J. Simon; J. O'Hara in *Ideal Knots*, eds. A. Stasiak , V. Katritch and L. H. Kauffman, (World Scientific, Singapore, 1998).

[11] P. Pierański, S. Kasas, G. Dietler, J. Dubochet and A. Stasiak, *Localization of breakage points in knotted strings*, submitted to New J. Phys. (2000).

[12] P. Pierański in *Ideal Knots*, eds. A. Stasiak, V. Katritch and L. H. Kauffman, (World Scientific, Singapore, 1998)

[13] D. Rolfsen D *Knots and Links* (Berkeley: Publish or Perish Press, 1976).

[14] C. Liang and K. Mislow, *J. Math. Chem.* **15**, 1 (1994).

[15] L. H. Kauffman, *Knots and Physics* (World Scientific Publishing Co., 1993) p.325.

[16] A. Stasiak, V. Katritch, J. Bednar, D. Michoud and J. Dubochet, *Nature* **384**, 122 (1996)

[17] V. Vologodskii, N. Crisona, B. Laurie, P. Pierański, V. Katritch, J. Dubochet and A. Stasiak, *J. Mol. Biol.* **278**, 1-3 (1998).